\documentclass[runningheads]{llncs}
\usepackage{rotating}
\usepackage[T1]{fontenc}
\usepackage{booktabs}
\usepackage{tabularx}
\usepackage{multirow}
\usepackage{graphicx}
\usepackage{amsfonts,amssymb,dsfont,amsmath}
\usepackage{xcolor}
\usepackage[misc]{ifsym}
\usepackage{tikz}
\usepackage{breakcites}
\usepackage[colorlinks=true,citecolor=blue,urlcolor=black]{hyperref}

\usepackage{color}

\usepackage{soul}

\newboolean{showcomments}
\setboolean{showcomments}{true}

\ifthenelse{\boolean{showcomments}}{
    \newcommand{\mynote}[3]{
        \fbox{\bfseries\sffamily\scriptsize#1:}
        {\small$\blacktriangleright$\textsf{\emph{\color{#3}{#2}}}$\blacktriangleleft$}
    }
}{
    \newcommand{\mynote}[3]{}
}

\begin{document}
\title{Universal Topology Refinement for Medical Image Segmentation with Polynomial Feature Synthesis}

\titlerunning{Universal Topology Refinement}
%
%

\author{Liu Li\inst{1}$^{(\textrm{\Letter})}$\and
Hanchun Wang\inst{1}\thanks{Liu Li and Hanchun Wang contributed equally to this work.}\and 
Matthew Baugh\inst{1}\and 
Qiang Ma\inst{1} \and 
Weitong Zhang\inst{1} \and
Cheng Ouyang\inst{2,1} \and
Daniel Rueckert \inst{1,3}\and
Bernhard Kainz\inst{1,4}
}
 
\authorrunning{L. Li et al.}
\institute{Imperial College London, UK \\\email{liu.li20@imperial.ac.uk} \and
University of Oxford, UK  \and
 Klinikum rechts der Isar, Technical University of Munich, Germany \and Friedrich-Alexander-Universit\"at Erlangen-N\"urnberg, Germany }

%
%
%

\maketitle              
\begin{abstract}
Although existing medical image segmentation methods provide impressive pixel-wise accuracy, they often neglect topological correctness, making their segmentations unusable for many downstream tasks. One option is to retrain such models whilst including a topology-driven loss component. However, this is computationally expensive and often impractical. A better solution would be to have a versatile plug-and-play topology refinement method that is compatible with any domain-specific segmentation pipeline.
Directly training a post-processing model to mitigate topological errors often fails as such models tend to be biased towards the topological errors of a target segmentation network. The diversity of these errors is confined to the information provided by a labelled training set, which is especially problematic for small datasets. 
Our method solves this problem by training a model-agnostic topology refinement network with synthetic segmentations that cover a wide variety of topological errors.
Inspired by the Stone-Weierstrass theorem, we synthesize topology-perturbation masks with randomly sampled coefficients of orthogonal polynomial bases, which ensures a complete and unbiased representation.
Practically, we verified the efficiency and effectiveness of our methods as being compatible with multiple families of polynomial bases, and show evidence that our universal plug-and-play topology refinement network outperforms both existing topology-driven learning-based and post-processing methods.
We also show that combining our method with learning-based models provides an effortless add-on, which can further improve the performance of existing approaches.

\end{abstract}
%
%
\section{Introduction}
\parskip0pt

The emergence of foundation models~\cite{zhou2023foundation} and SAM models~\cite{ma2024segment,mazurowski2023segment} has led to significant progress in medical image segmentation.
However, these methods are first and foremost designed to achieve high pixel-wise accuracy and neglect the topological correctness of the segmentation.
Topological correctness is often vital in medical imaging because many downstream tasks, such as vessel analysis~\cite{yao2023tag} or cortical surface analysis~\cite{dale1999cortical}, depend on a correct topology.

To address this challenge, several topology-aware loss functions have been designed to force deep learning models to pay more attention to the topological correctness of the segmentation~\cite{hu2019topology,clough2020topological,clough2019explicit,stucki2023topologically,berger2024topologically}. Although these methods can improve topological correctness during training by enforcing priors derived from algebraic topology, their computational complexity increases exponentially as a function of the input dimensionality~\cite{hu2022structure}, which further hinder their ability to employ in foundation models such as SAM models. 
Also, as these approaches still infer the topology information from the input image, they are susceptible to image artifacts, such as noise, partial volume effects or low image resolution, limiting the effectiveness of their topological priors. 

An alternative approach for topology correction utilizes a post-processing network to learn the topology priors directly from the predictions of a segmentation network~\cite{li2023robust,larrazabal2020post,gondara2016medical}.
By operating on the segmentations themselves these methods avoid the issues arising from learning from noisy images.
Furthermore, topology refinement during post-processing also provides an approach in which already pre-trained segmentation networks can be used. 

\begin{figure}[t]
\centering
    \centering
    \includegraphics[width=1.0\linewidth]{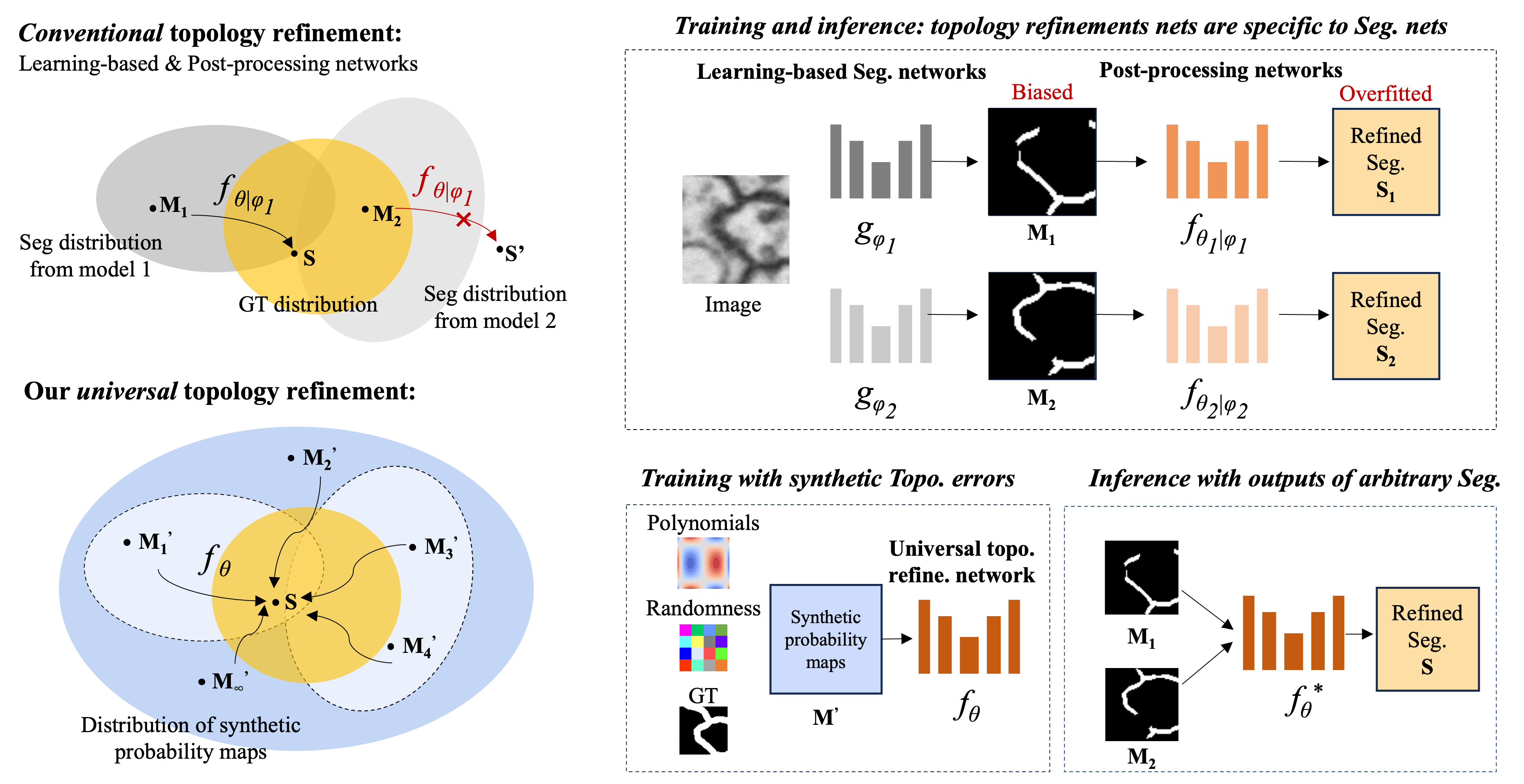}
    \caption{\textbf{Top row:} 
    Conventional topology-refinement post-processing networks are trained on, and biased towards the topological errors of a specific upstream segmentation network $g_\phi$, leading to errors when applied to a different $g_\phi$. Re-training is required when $g_\phi$ changes.
    \textbf{Bottom row:} To train an unbiased topology-refinement network $f_{\theta}$, our method synthesizes training samples from an unbiased distribution, as illustrated in the blue ellipses. This covers plausible real segmentations with different topological errors, generated by complete and orthogonal polynomials. During inference, segmentations $\mathbf{M}$ from $g_\phi$ can be refined by a universally trained $f_{\theta}$ network.
}
\label{fig:fig1}
\end{figure}

However, existing topology refinement networks $f_{\theta|\phi}$ are biased towards the output probability distribution of the specific segmentation network $g_\phi$ it is trained to complement~\cite{li2023robust}. 
This approach leads to poor generalization when applying a refinement network to new segmentation networks.
Fig.~\ref{fig:fig1} illustrates this with different $g_\phi$'s, each requiring a separate $f_{\theta|\phi}$ to be trained. 
In addition, for small datasets, model-specific post-processing may overfit to the individual topology errors in the training set causing it to fail even when paired with its corresponding $g_\phi$. 


In this paper, we propose a universal post-processing topology refinement network that can be plugged into 
any existing deep learning segmentation networks without $g_\phi$-specific tuning. 
To train this network, we propose to synthesize training sets that cover a large range of possible topological errors, unbiased towards any specific $g_\phi$. 
Notably, to ensure the diversity of synthesized errors in terms of size, location, and severity, we draw topology-perturbation masks from random linear combinations of continuous orthogonal polynomial bases.
This provides the following advantages: 1. The location, size, and number of topological errors can be controlled with the orders and coefficients of the polynomials~\cite{funaro2008polynomial,sw_theorem}. 2. Multiplying the resulting continuous perturbation masks with the ground truth (GT) segmentation produces plausible topological errors that reflect those exhibited in the real world, \emph{i.e.} they manifest as gradual, smooth deviations in predictive probabilities rather than sharp discontinuities \cite{clough2019explicit,clough2020topological}. 
Our code is publicly available here$\footnote{\scriptsize \href{https://github.com/smilell/Universal-Topology-Refinement}{https://github.com/smilell/Universal-Topology-Refinement}}$. 



\noindent\textbf{Contributions.} Our post-processing network is exclusively trained using unbiased synthetic segmentation probability maps, which are derived from high-dimensional polynomials.
When applied during inference, our model is capable of refining the probability maps produced by any pre-existing segmentation networks. Hence, in summary  
\textbf{(1)} We propose a universal topology refinement pipeline to revise neural-network-derived segmentations. Our approach, which serves as a plug-and-play module, can adapt to any segmentation backbone. 
\textbf{(2)} To address a wide range of potential topological errors, we develop two types of continuous topology-perturbation synthesis and 
segmentation map synthesis. These designs guarantee the variety and completeness of the synthetic training data.
\textbf{(3)} By leveraging an abundance of synthetic samples our approach exhibits more robust performance, especially for high-dimensional datasets with limited GT labels. We demonstrate superior performance compared to existing state-of-the-art methods on the 3D CoW dataset~\cite{yang2023benchmarking}.
Additionally, we showcase our method's capacity for generalization by revealing its potential to improve existing segmentation models without the need for fine-tuning.


\noindent\textbf{Related work.}
\textit{Learning-based} methods with explicit constraints introduce additional loss terms derived from either shape~\cite{shit2021cldice,kervadec2019boundary} or topology priors~\cite{hu2019topology,clough2020topological,clough2019explicit,li2022fetal,de2021segmentation,byrne2022persistent,stucki2022topologically,hu2022structure,hu2021topology}. 
Specifically, \cite{shit2021cldice} and \cite{kervadec2019boundary} compare the center lines and boundaries between the prediction and GT, introducing strong shape priors into the network. For topology-aware constraints, persistent homology (PH)~\cite{hu2019topology,clough2020topological,clough2019explicit,stucki2023topologically}, Morse theory~\cite{hu2022learning,hu2021topology} and homotopy warping~\cite{hu2022structure} have been studied to \emph{guarantee} topological correctness. 
However, these approaches require careful balancing between pixel-wise and topology-aware loss terms. 

\textit{Post-processing} topology refinement refines the topology by using
heuristics or training another network on an initial segmentation prediction with a GT as supervision~\cite{larrazabal2020post,gondara2016medical,li2023robust}. A topology-aware post-processing network has been first proposed in~\cite{li2023robust} to identify disrupted topology regions based on Euler characteristics. However, these post-processing methods are model-specific and exhibit bias toward their training data, limiting their ability to generalize effectively to scenarios with different topological errors.








\section{Method}

\begin{figure}[t]
\vspace{-0.5cm}
\centering
    \centering
    \includegraphics[width=1\linewidth]{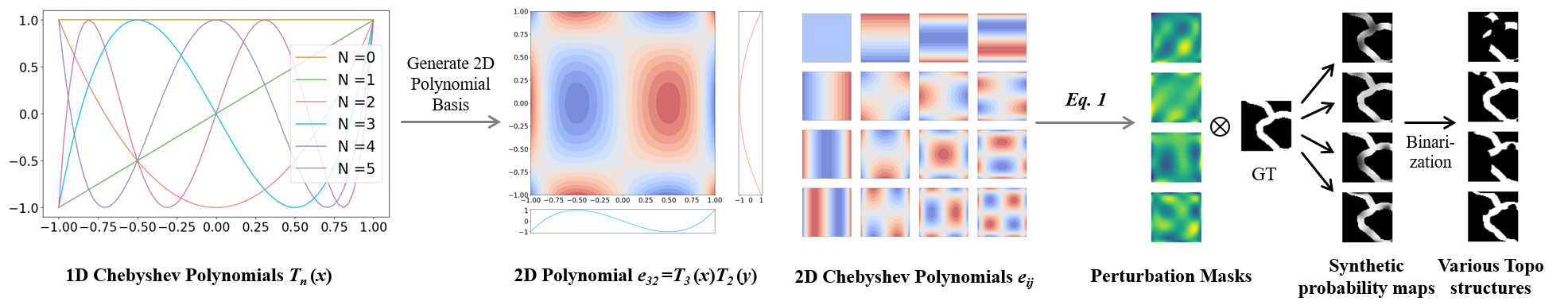}
    \caption{Workflow for generating arbitrary 2D topology error maps from orthogonal polynomials (Chebyshev polynomials as examples).}
\label{fig:fig2}
\vspace{-0.5cm}
\end{figure}

To train a model-agnostic post-processing network, we synthesize unbiased topology-aware segmentation maps that can cover a diverse range of possible cases for topological errors. 
We show that complete and orthogonal polynomials are ideal candidates for synthesizing various and unbiased topology-perturbation masks. 

\noindent\textbf{Preliminary on orthogonal polynomials.}
Consider a continuous function $f: [0,1] \to [0,1]$ on the interval, $f$ can be approximated arbitrarily close by polynomials under the weighted $l^2$ norm by the Stone–Weierstrass approximation theorem~\cite{sw_theorem}. Therefore, it is natural to choose a polynomial basis $\{\phi_n\}_{n=0}^\infty$ and use a linear combination of basis functions $p(x)=\sum_{n=0}^\infty a_n\phi_n(x)$ to represent a segmentation function $f(x)$ completely.


\begin{table}[!h]
\centering
\label{tab:polys}
\caption{Summary of Legendre, Hermite-Gaussian, and Chebyshev Polynomials and their orthogonality intervals.}
\begin{tabular}{p{4.2cm}cp{5.2cm}cc}
\toprule
Polynomial & &  Definition & & Interval\\  %
\midrule
Legendre $P_n$ & & $P_n(x) = (2^n n!)^{-1} \frac{d^n}{dx^n}(x^2-1)^n$& & $[-1, 1]$ \\   %

Chebyshev (1st kind) $T_n$ & & $T_n(x) = \cos(n \arccos(x))$ & &$[-1, 1]$\\  

Hermite-Gaussian $\psi_n$ & & $\psi_n(x)=\left(2^n n ! \sqrt{\pi}\right)^{-\frac{1}{2}} e^{-\frac{x^2}{3}} H_n(x)$ & &$(-\infty, \infty)$\\     
\bottomrule
\end{tabular}

\label{tab:polynomials}
\end{table}


Practically, to represent the topology perturbation mask, we use the sum of the first $N$ basis $p'(x)=\sum_{n=0}^N a_n\phi_n(x)$ of orthogonal polynomial basis. The theoretical insight of choosing an orthogonal basis is proved in the Gram–Schmidt process~\cite{chebyshev_poly}. 
Notably, the orthogonality of the basis ensures the location of potentially induced topological errors to be spatially uniform (therefore unbiased), when the coefficients are sampled from Gaussian distributions~\cite{sampling}. 
We experiment with three commonly used orthogonal polynomial bases in Tab.~\ref{tab:polynomials} based on good analytical properties discussed in the Appendix. 


\noindent\textbf{Topology-perturbation mask generation.}
A high-dimensional orthogonal polynomial basis $\Phi_{3D}=\{e_{ijk}\}$ (3D for example) can be constructed from a 1D basis $\Phi=\{\phi_n\}_{n=0}^\infty$ by tensor products $\Phi_{3D}=\Phi \otimes \Phi \otimes \Phi$ such that $e_{ijk}(x,y,z)=\phi_i(x)\phi_j(y)\phi_k(z)$.  Fig.~\ref{fig:fig2} illustrates an example of a derived 2D basis $e_{32}(x,y)=T_3(x)T_2(y)$, and its concurrent 2D basis generated by Chebyshev polynomials up to $4^{th}$ orders.

To generate a 3D continuous function $h(x,y,z)$, a polynomial basis set $\{\phi_n\}_{n=0}^N$ is selected to create 3D basis $\{e_{ijk}\}$. By randomly sampling the coefficients $a_{ijk}$ from a Gaussian distribution, we can sample different perturbation masks with normalized spatial coordinates $x, y, z\in[-1,1]$:
\begin{equation}
h(x,y,z) = \sum_{i,j,k=1}^{N} a_{ijk} e_{ijk}(x,y,z), a_{ijk}\sim \mathcal{N}(0, 1),
\label{eq1}
\end{equation}
where $(i,j,k)$ are indices for 1-D polynomials, and $N$ is the number of chosen 3D basis. In principle, a larger $N$ allows the mask to have higher spatial frequencies and introduce more critical points under which the topology of GT may be distorted (Appendix).

An important advantage of such polynomial basis lies in their capacity to control topological changes due to their analytical properties. The indicative of topological features, \emph{e.g.} Betti numbers and persistent diagram, can be determined through the critical points on the Morse function, which is derived from the probability map~\cite{hu2021topology, hu2022learning}. By multiplying a topology-perturbation mask to GT, the critical points of the Morse function undergo modifications, which are aware of and dependent on the preceding Morse function and its following topological features. 

\noindent\textbf{Synthesizing segmentations with topological errors.}
The obtained topology perturbation masks are multiplied with ground truth segmentations, yielding the training data for the proposed universal topology refinement network $f_{\theta}$. 

Specifically, the perturbation masks $\mathbf{H}$ (with its elements $H_{m,n,l}$ at coordinate $(m,n,l)$) are discretized from the continuous representation $h(x,y,z)$ over a Cartesian grid at resolution $[N_x, N_y, N_z]$:

\begin{equation}
H_{m,n,l} = h(x_m, y_n, z_l) = h\left(\frac{m}{N_x }, \frac{n}{N_y}, \frac{l}{N_z}\right).
\label{eq2}
\end{equation}

Given the normalized topology-perturbation mask $\mathbf{H}$, as shown in Fig.~\ref{fig:fig2} in green, the segmentation probability maps $\mathbf{M}'$ are generated by using the element-wise multiplication between $\mathbf{H}$ and GT. Note that our topology-aware method does not extract the GT topology explicitly, instead, by perturbing the GT with mask $\mathbf{H}$, 
we simulate incorrect, continuous softmax segmentation predictions that lead to topology errors. The change to topology can be visualized by thresholding and evaluated by topological features such as Betti number or persistent diagram. 
In Fig.~\ref{fig:fig2} we further show the binarization results of $\mathbf{M}'$, as an example to illustrate the change of topology in terms of Betti number.


In addition to the topological disruptions systematically modeled by complete and orthogonal polynomials, we further simulate errors caused by over-segmentation. Superfluous segmentation components are modeled by perturbing $\mathbf{M}'$ with multiplicative Gaussian noise. This enhances the robustness of $f_{\theta}$.

\noindent\textbf{Training and inference.} The generated segmentation probability maps are used to train $f_\theta$. During inference, the trained $f_\theta$ is fixed and applied to segmentation probability maps from existing segmentation models. Note that our model is adaptable to any existing domain-specific segmentation network, which is demonstrated in the Experiment section.


\section{Experiments}
\noindent\textbf{Datasets.} We conduct experiments on the TopCoW dataset which contains magnetic resonance angiography (MRA) data~\cite{yang2023benchmarking}. TopCoW is the first 3D dataset with voxel-level annotations for the vessels of the Circle of Willis (CoW). 
We randomly split the 90 publicly available labeled volumes into 55 for training, 5 for validation, and 30 for testing. 


\noindent\textbf{Hyper-parameters.} We conducted experiments on three polynomial families: Legendre, Hermite-Gaussian, and Chebyshev, with the polynomials up $N=10$. 
$[N_x, N_y, N_z]$ in Eq.~\ref{eq2} are set to $[64, 64, 64]$ to inject localized corruptions. 

\noindent\textbf{Implementation.} All the experiments are conducted on one NVIDIA RTX 3080 GPU. Our method randomly generates synthetic data in real-time, with a training time of 0.29 seconds per batch, 21.0 times faster than the baseline method PH-loss~\cite{hu2019topology}.

\noindent\textbf{Evaluation metrics.} The topological performance is evaluated by Betti errors $e_i = \mid \beta^{\mathrm{pred}}_{i} - \beta^{\mathrm{gt}}_{i}\mid$, where $i\in \{0,1\} $ is the dimension. We also report the mean Betti error as $e = e_0 + e_1$. Segmentation performance is evaluated by Dice score.

\begin{table}[t]\setlength{\tabcolsep}{4pt}
  \centering
  \caption{Topological quality metrics for our method on the 3D TopCoW dataset compared with two baseline methods: learning-based and post-processing. }
    \begin{tabular}{llccccc}
           \multicolumn{1}{l}{} & \multicolumn{1}{l}{Method} & Dice$\uparrow$ & 
          & Betti error$\downarrow$ & Betti 0$\downarrow$ & Betti 1$\downarrow$ \\
    \midrule
 & 
    cl-Dice loss~\cite{shit2021cldice} & $94.48_{\pm2.74}$ &       & $2.50_{\pm1.54}$ & $1.43_{\pm1.05}$ & $1.07_{\pm1.44}$ \\
             Learning  & Boundary loss~\cite{kervadec2019boundary} & $94.55_{\pm2.62}$ &       & $2.57_{\pm1.56}$ &  $1.60_{\pm1.20}$  &  $0.97_{\pm1.40}$\\
          -based & PH loss~\cite{hu2019topology} & $94.62_{\pm2.41}$ &       & $2.60_{\pm1.70}$ & $1.60_{\pm1.40}$   & $1.00_{\pm1.46}$ \\
               & Warp loss~\cite{hu2022structure} & $93.99_{\pm3.23}$ &       & $2.90_{\pm2.07}$ & $1.87_{\pm1.45}$ & $1.03_{\pm1.49}$\\
\cmidrule{1-7}          
         Post- & Hand-crafted~\cite{vlachos2010multi,soomro2022image} & $94.29_{\pm3.11}$ &  & $2.43_{\pm2.01}$  & $1.40_{\pm1.25}$  & $1.03_{\pm1.78}$ \\
            processing    & DAE~\cite{gondara2016medical}   & $94.61_{\pm2.54}$ &       & $2.70_{\pm1.83}$ & $1.57_{\pm1.28}$ & $1.13_{\pm1.71}$ \\
\cmidrule{1-7}         
            Universal  & \textbf{Ours}  & $94.26_{\pm3.08}$ &       & $\mathbf{1.97}_{\pm1.52}$ & $\mathbf{1.07}_{\pm0.91}$ & $\mathbf{0.90}_{\pm1.48}$\\
    \end{tabular}%
  \label{tab:main}%
\end{table}%


\noindent\textbf{Quantitative evaluation.} 
We compare the topology preservation and refinement abilities of our methods with four learning-based shape-/topology-driven baselines: cl-Dice loss~\cite{shit2021cldice}, boundary loss~\cite{kervadec2019boundary}, PH loss~\cite{hu2019topology}, warp loss~\cite{hu2022structure}, and two post-processing methods: hand-crafted noise filter~\cite{vlachos2010multi,soomro2022image}, and denoising autoencoder (DAE)~\cite{gondara2016medical}.  Our approach yields the best performance in terms of topological quality.
For a fair comparison, U-Net~\cite{ronneberger2015u,isensee2018nnu} is employed as the segmentation and post-processing backbone with default cross-entropy loss. 
For learning-based methods, additional loss terms are incorporated during training with the provided weights from the corresponding papers. Note that post-processing methods are tested based on the baseline predictions from the U-Net trained with default cross-entropy loss.

\begin{figure}[t]
\centering
\includegraphics[width=1.0\linewidth]{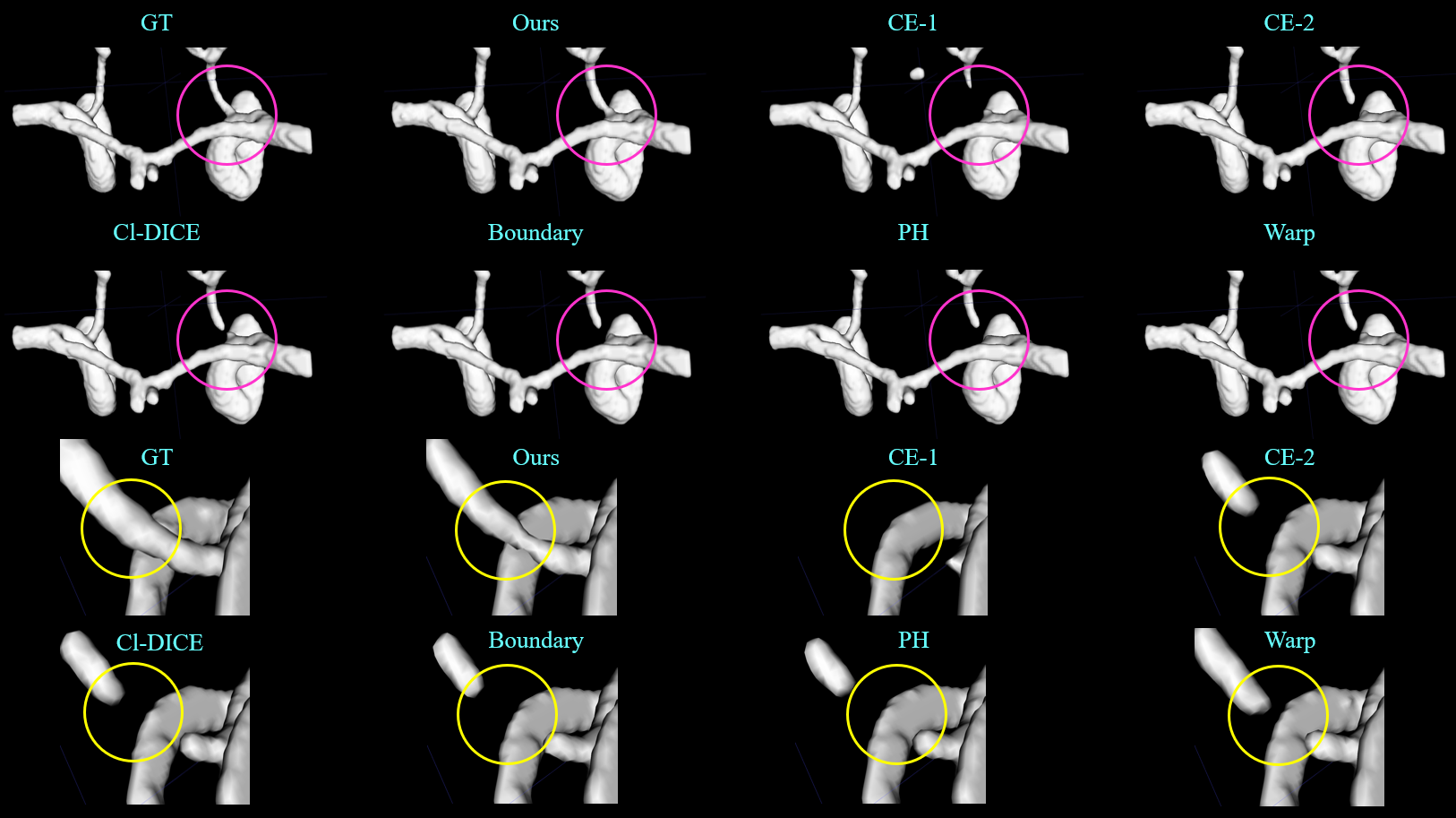}  
    \caption{Qualitative results for topology preservation/refinement. Our approach successfully revises disconnected structures that should be connected. }
\label{fig:qualititive}
\end{figure}

As shown in Tab.~\ref{tab:main}, despite the similar pixel-wise accuracy in terms of Dice score across all methods, our method achieves significant improvements in terms of Betti errors. 
We have performed Wilcoxon signed-rank tests~\cite{wilcoxon1992individual} to demonstrate that our methods outperform other baseline approaches in terms of Betti errors with p-values of 0.020, 0.017, 0.012, 0.002, 0.041, and 0.002, respectively (all $<0.05$). Qualitative examples are shown in Fig.~\ref{fig:qualititive}.

We further demonstrate the generalization capability of our methods by applying our topology refinement network trained purely with synthetic data to different segmentation models. We compare our performance with two existing post-processing methods: model-specific DAE~\cite{gondara2016medical} and hand-crafted noise filtering~\cite{soomro2022image,vlachos2010multi}. Table~\ref{tab:second} illustrates the universal efficacy of our method in improving topological correctness, agnostic to the output distribution shifts of the pre-trained segmentation models.

It is worth mentioning that our model, trained on synthetic data using only the GT labelmap, is agnostic to covariate (domain) shifts, e.g., imaging devices and scanning protocols. Such a design makes our method a robust plug-and-play option for various upstream segmentation models. Retraining our model on new tasks with different topological structures is straightforward and quick.

\begin{table}[t]
  \centering
  \caption{Comparison of post-processing methods plugged into five raw learning-based methods. Our method can universally improve the topological performance of different methods, outperforming the performance of existing post-processing methods.
  }
  \resizebox{1.0\columnwidth}{!}{%
    \begin{tabular}{lp{0.1cm}ccccp{0.1cm}ccccp{0.1cm}ccccp{0.1cm}cccc}
    \multicolumn{1}{c}{\textbf{\textit{}}} && \multicolumn{4}{c}{Raw}      && \multicolumn{4}{c}{Hand-crafted~\cite{soomro2022image}} && \multicolumn{4}{c}{Model-specific~\cite{gondara2016medical}} && \multicolumn{4}{c}{Ours} \\
    \cmidrule{3-6}\cmidrule{8-21}
    \multicolumn{1}{c}{\textbf{\textit{}}} && Dice$\uparrow$  & $e$$\downarrow$ & $e_0$ $\downarrow$ & $e_1\downarrow$ && Dice$\uparrow$ & $e$$\downarrow$ & $e_0$$\downarrow$ & $e_1$$\downarrow$ && Dice$\uparrow$  & $e$$\downarrow$ & $e_0$$\downarrow$ & $e_1$$\downarrow$ && Dice$\uparrow$  & $e$$\downarrow$ & $e_0$$\downarrow$ & $e_1$$\downarrow$  \\
    \midrule
    CE~\cite{ronneberger2015u}  & & 94.45  & 3.00  & 1.93  & 1.07  && 94.29  & 2.43  & 1.40  & 1.03  && 94.61  & 2.70  & 1.57  & 1.13  && 94.26  & \textbf{1.97} & \textbf{1.07} & \textbf{0.90} \\
    cl-Dice~\cite{shit2021cldice} && 94.48  & 2.50  & 1.43  & 1.07  && 94.49  & 2.20  & 1.13  & 1.07 & & 94.52  & 2.10  & \textbf{1.17} & 0.93  && 94.16  & \textbf{2.00} & \textbf{1.17} & \textbf{0.83} \\
    Boundary~\cite{kervadec2019boundary} && 94.55  & 2.57  & 1.60  & 0.97  && 94.56  & 2.23  & 1.27 & 0.97  && 94.54  & 2.27  & 1.37  & 0.90 & & 94.06  & \textbf{2.07} & \textbf{1.20} & \textbf{0.87} \\
    PH~\cite{hu2019topology} && 94.62  & 2.60  & 1.60  & 1.00  && 94.62  & 2.27  & 1.27  & 1.00  && 94.55  & 2.63  & 1.67  & \textbf{0.97} && 94.31  & \textbf{2.00} & \textbf{1.03} & \textbf{0.97} \\
    Warp~\cite{hu2022structure} && 93.99  & 2.90  & 1.87  & 1.03  && 93.99  & 2.27  & 1.30  & 0.97  && 94.37  & 2.27  & 1.33  & 0.93  && 93.44  & \textbf{1.83} & \textbf{0.97} & \textbf{0.87} \\
    \end{tabular}%
    }
  \label{tab:second}%
\end{table}%


\begin{figure}[t]
\centering
\includegraphics[width=1.0\linewidth] 
 {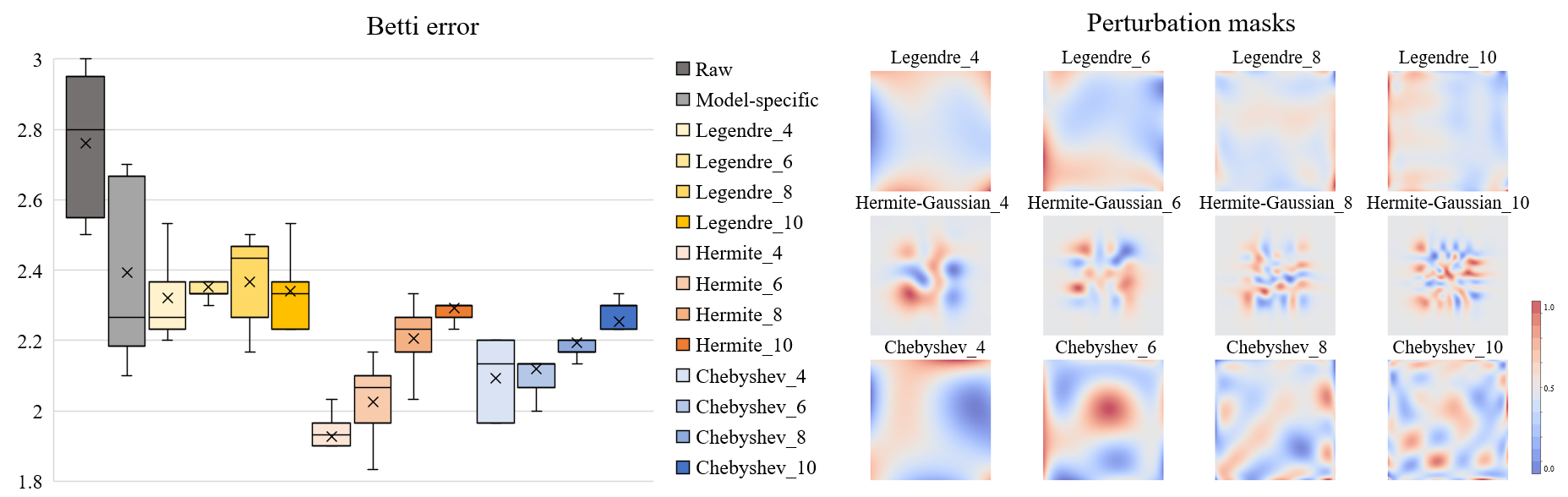}  
    \caption{Ablation study for different types of polynomials. Left: Betti error across models trained with different polynomials. Right: Perturbation masks generated from Legendre, Hermite-Gaussian, and Chebyshev polynomials at order 4, 6, 8, and 10. }
\label{fig:ablation}
\end{figure}

\noindent\textbf{Ablation study.}\label{ablation}
We explore the impact of polynomial families across orders. In Fig.~\ref{fig:ablation}-left, each model is trained with segmentations modulated by different types of perturbations. For a comprehensive evaluation, each model is tested by the output predictions from all five segmentation models~\cite{ronneberger2015u,shit2021cldice,kervadec2019boundary,hu2019topology,hu2022structure}. 
Our method surpasses raw segmentation and conventional post-processing in mean Betti error. Legendre polynomial performance remains consistent across orders, while Hermite-Gaussian and Chebyshev performance increases with decreasing order due to unrealistic features, as seen in Fig.~\ref{fig:ablation}-right. As shown, lower-order masks exhibit smoother gradients with fewer saddle points, whereas higher-order masks incorporate more high-frequency terms, thus generating unrealistic masks.

\noindent\textbf{Discussion.}
We first demonstrate the effectiveness of our method by applying it to the segmentations from the default U-Net trained with cross-entropy loss. Compared to existing methods, ours achieves the best topological performance. 
In addition, we demonstrate that our method is a light plug-and-play topology refinement method compatible with any domain-specific segmentation pipelines. We plug our method into five learning-based methods and show our method can universally improve the topological performance. 
We also conduct the ablation study on multiple families of polynomial basis and reached the conclusion that polynomials can consistently improve topological performance. The best setting is the synthesis samples with a lower order of polynomials since the actual probability map is mainly constructed by low-frequency components.

\section{Conclusion}
Although image segmentation is often considered a solved problem, challenges persist in achieving topological accuracy to the level required by downstream tasks.
This paper introduces a universal topology refinement model trained using polynomial-based data synthesis.
Our method can be applied as a plug-and-play module to existing segmentation methods and achieve state-of-the-art performance without requiring model-specific fine-tuning.
Future research can explore the trade-off between the topological errors caused by over-segmentation and under-segmentation, and investigate the sampling of polynomial coefficients for improved performance. 
One potential direction is employing adversarial learning to investigate the best settings.

\noindent\textbf{Acknowledgements:} 
This project is supported by Lee Family Scholarship from ICL. HPC resources are provided by the Erlangen National High Performance Computing Center (NHR@FAU) of the Friedrich-Alexander-Universität Erlangen-Nürnberg (FAU) under the NHR project b143dc and b180dc. NHR funding is provided by federal and Bavarian state authorities. NHR@FAU hardware is partially funded by the German Research Foundation (DFG) – 440719683. Support was also received by the ERC - projects MIA-NORMAL 101083647 and Deep4MI 884622 as well as DFG KA 5801/2-1, INST 90/1351-1.

\subsubsection{\discintname}
The authors have no competing interests to declare that are relevant to the content of this article.


\bibliographystyle{splncs04}
\bibliography{mybib}

\begin{thebibliography}{10}
\providecommand{\url}[1]{\texttt{#1}}
\providecommand{\urlprefix}{URL }
\providecommand{\doi}[1]{https://doi.org/#1}

\bibitem{berger2024topologically}
Berger, A.H., Stucki, N., Lux, L., Buergin, V., Shit, S., Banaszak, A.,
  Rueckert, D., Bauer, U., Paetzold, J.C.: Topologically faithful multi-class
  segmentation in medical images. arXiv preprint arXiv:2403.11001  (2024)

\bibitem{byrne2022persistent}
Byrne, N., Clough, J.R., Valverde, I., Montana, G., King, A.P.: A persistent
  homology-based topological loss for cnn-based multiclass segmentation of cmr.
  IEEE Transactions on Medical Imaging  \textbf{42}(1),  3--14 (2022)

\bibitem{clough2020topological}
Clough, J.R., Byrne, N., Oksuz, I., Zimmer, V.A., Schnabel, J.A., King, A.P.: A
  topological loss function for deep-learning based image segmentation using
  persistent homology. IEEE Transactions on Pattern Analysis and Machine
  Intelligence  \textbf{44}(12),  8766--8778 (2020)

\bibitem{clough2019explicit}
Clough, J.R., Oksuz, I., Byrne, N., Schnabel, J.A., King, A.P.: Explicit
  topological priors for deep-learning based image segmentation using
  persistent homology. In: Information Processing in Medical Imaging: 26th
  International Conference, IPMI 2019, Hong Kong, China, June 2--7, 2019,
  Proceedings 26. pp. 16--28. Springer (2019)

\bibitem{dale1999cortical}
Dale, A.M., Fischl, B., Sereno, M.I.: Cortical surface-based analysis: {I}.
  segmentation and surface reconstruction. Neuroimage  \textbf{9}(2),  179--194
  (1999)

\bibitem{de2021segmentation}
de~Dumast, P., Kebiri, H., Atat, C., Dunet, V., Koob, M., Cuadra, M.B.:
  Segmentation of the cortical plate in fetal brain mri with a topological
  loss. In: Uncertainty for Safe Utilization of Machine Learning in Medical
  Imaging, and Perinatal Imaging, Placental and Preterm Image Analysis: 3rd
  International Workshop, UNSURE 2021, and 6th International Workshop, PIPPI
  2021, Held in Conjunction with MICCAI 2021, Strasbourg, France, October 1,
  2021, Proceedings 3. pp. 200--209. Springer (2021)

\bibitem{funaro2008polynomial}
Funaro, D.: Polynomial approximation of differential equations, vol.~8.
  Springer Science \& Business Media (2008)

\bibitem{sampling}
García~García, A.: Orthogonal sampling formulas: A unified approach. Society
  for Industrial and Applied Mathematics  \textbf{42},  499--512 (09 2000).
  \doi{10.1137/S0036144599363497}

\bibitem{gondara2016medical}
Gondara, L.: Medical image denoising using convolutional denoising
  autoencoders. In: 2016 IEEE 16th international conference on data mining
  workshops (ICDMW). pp. 241--246. IEEE (2016)

\bibitem{hu2022structure}
Hu, X.: Structure-aware image segmentation with homotopy warping. Advances in
  Neural Information Processing Systems  \textbf{35},  24046--24059 (2022)

\bibitem{hu2019topology}
Hu, X., Li, F., Samaras, D., Chen, C.: Topology-preserving deep image
  segmentation. Advances in neural information processing systems  \textbf{32}
  (2019)

\bibitem{hu2022learning}
Hu, X., Samaras, D., Chen, C.: Learning probabilistic topological
  representations using discrete morse theory (2022)

\bibitem{hu2021topology}
Hu, X., Wang, Y., Fuxin, L., Samaras, D., Chen, C.: Topology-aware segmentation
  using discrete morse theory. arXiv preprint arXiv:2103.09992  (2021)

\bibitem{isensee2018nnu}
Isensee, F., Petersen, J., Klein, A., Zimmerer, D., Jaeger, P.F., Kohl, S.,
  Wasserthal, J., Koehler, G., Norajitra, T., Wirkert, S., et~al.: nnu-net:
  Self-adapting framework for u-net-based medical image segmentation.
  arXiv:1809.10486  (2018)

\bibitem{kervadec2019boundary}
Kervadec, H., Bouchtiba, J., Desrosiers, C., Granger, E., Dolz, J., Ayed, I.B.:
  Boundary loss for highly unbalanced segmentation. In: International
  conference on medical imaging with deep learning. pp. 285--296. PMLR (2019)

\bibitem{larrazabal2020post}
Larrazabal, A.J., Mart{\'\i}nez, C., Glocker, B., Ferrante, E.: Post-dae:
  anatomically plausible segmentation via post-processing with denoising
  autoencoders. IEEE Transactions on Medical Imaging  \textbf{39}(12),
  3813--3820 (2020)

\bibitem{li2022fetal}
Li, L., Ma, Q., Li, Z., Ouyang, C., Zhang, W., Price, A., Kyriakopoulou, V.,
  Grande, L.C., Makropoulos, A., Hajnal, J., et~al.: Fetal cortex segmentation
  with topology and thickness loss constraints. In: Ethical and Philosophical
  Issues in Medical Imaging, Multimodal Learning and Fusion Across Scales for
  Clinical Decision Support, and Topological Data Analysis for Biomedical
  Imaging: 1st International Workshop, EPIMI 2022, 12th International Workshop,
  ML-CDS 2022, 2nd International Workshop, TDA4BiomedicalImaging, Held in
  Conjunction with MICCAI 2022, Singapore, September 18--22, 2022, Proceedings.
  pp. 123--133. Springer (2022)

\bibitem{li2023robust}
Li, L., Ma, Q., Ouyang, C., Li, Z., Meng, Q., Zhang, W., Qiao, M.,
  Kyriakopoulou, V., Hajnal, J.V., Rueckert, D., et~al.: Robust segmentation
  via topology violation detection and feature synthesis. In: International
  Conference on Medical Image Computing and Computer-Assisted Intervention. pp.
  67--77. Springer (2023)

\bibitem{ma2024segment}
Ma, J., He, Y., Li, F., Han, L., You, C., Wang, B.: Segment anything in medical
  images. Nature Communications  \textbf{15}(1), ~654 (2024)

\bibitem{mazurowski2023segment}
Mazurowski, M.A., Dong, H., Gu, H., Yang, J., Konz, N., Zhang, Y.: Segment
  anything model for medical image analysis: an experimental study. Medical
  Image Analysis  \textbf{89},  102918 (2023)

\bibitem{chebyshev_poly}
Onderzoek, B.: Chebyshev approximation. Ph.D. thesis, University of Groningen
  (2017)

\bibitem{ronneberger2015u}
Ronneberger, O., Fischer, P., Brox, T.: U-net: Convolutional networks for
  biomedical image segmentation. In: Medical Image Computing and
  Computer-Assisted Intervention--MICCAI 2015: 18th International Conference,
  Munich, Germany, October 5-9, 2015, Proceedings, Part III 18. pp. 234--241.
  Springer (2015)

\bibitem{shit2021cldice}
Shit, S., Paetzold, J.C., Sekuboyina, A., Ezhov, I., Unger, A., Zhylka, A.,
  Pluim, J.P., Bauer, U., Menze, B.H.: cldice-a novel topology-preserving loss
  function for tubular structure segmentation. In: Proceedings of the IEEE/CVF
  Conference on Computer Vision and Pattern Recognition. pp. 16560--16569
  (2021)

\bibitem{soomro2022image}
Soomro, T.A., Zheng, L., Afifi, A.J., Ali, A., Soomro, S., Yin, M., Gao, J.:
  Image segmentation for mr brain tumor detection using machine learning: A
  review. IEEE Reviews in Biomedical Engineering  (2022)

\bibitem{sw_theorem}
Stone, M.H.: The generalized weierstrass approximation theorem. Mathematics
  Magazine  \textbf{21}(5),  237--254 (1948)

\bibitem{stucki2022topologically}
Stucki, N., Paetzold, J.C., Shit, S., Menze, B., Bauer, U.: Topologically
  faithful image segmentation via induced matching of persistence barcodes.
  arXiv preprint arXiv:2211.15272  (2022)

\bibitem{stucki2023topologically}
Stucki, N., Paetzold, J.C., Shit, S., Menze, B., Bauer, U.: Topologically
  faithful image segmentation via induced matching of persistence barcodes. In:
  International Conference on Machine Learning. pp. 32698--32727. PMLR (2023)

\bibitem{vlachos2010multi}
Vlachos, M., Dermatas, E.: Multi-scale retinal vessel segmentation using line
  tracking. Computerized Medical Imaging and Graphics  \textbf{34}(3),
  213--227 (2010)

\bibitem{wilcoxon1992individual}
Wilcoxon, F.: Individual comparisons by ranking methods. In: Breakthroughs in
  Statistics: Methodology and Distribution, pp. 196--202. Springer (1992)

\bibitem{yang2023benchmarking}
Yang, K., Musio, F., Ma, Y., Juchler, N., Paetzold, J.C., Al-Maskari, R.,
  H{\"o}her, L., Li, H.B., Hamamci, I.E., Sekuboyina, A., et~al.: Benchmarking
  the cow with the topcow challenge: Topology-aware anatomical segmentation of
  the circle of willis for cta and mra. arXiv preprint arXiv:2312.17670  (2023)

\bibitem{yao2023tag}
Yao, L., Shi, F., Wang, S., Zhang, X., Xue, Z., Cao, X., Zhan, Y., Chen, L.,
  Chen, Y., Song, B., et~al.: Tag-net: Topology-aware graph network for
  centerline-based vessel labeling. IEEE Transactions on Medical Imaging
  (2023)

\bibitem{zhou2023foundation}
Zhou, Y., Chia, M.A., Wagner, S.K., Ayhan, M.S., Williamson, D.J., Struyven,
  R.R., Liu, T., Xu, M., Lozano, M.G., Woodward-Court, P., et~al.: A foundation
  model for generalizable disease detection from retinal images. Nature
  \textbf{622}(7981),  156--163 (2023)

\end{thebibliography}

\end{document}